\documentclass[conference]{IEEEtran}
\usepackage{amsmath}
\usepackage{booktabs}
\usepackage{tabularx}
\usepackage{array}
\usepackage{graphicx}
\usepackage{svg}
\usepackage{xcolor}
\usepackage{subcaption}
\usepackage{comment}
\ifCLASSINFOpdf
\else
\fi
\AtBeginDocument{%
  \setlength{\abovedisplayskip}{3pt plus 1pt minus 1pt}%
  \setlength{\belowdisplayskip}{3pt plus 1pt minus 1pt}%
  \setlength{\abovedisplayshortskip}{1pt plus 1pt}%
  \setlength{\belowdisplayshortskip}{2pt plus 1pt}%
}
\makeatletter
\def\section{\@startsection{section}{1}{\z@}{1.1ex plus 0.5ex minus 0.2ex}%
{0.4ex plus 0.3ex minus 0ex}{\normalfont\normalsize\centering\scshape}}%
\def\subsection{\@startsection{subsection}{2}{\z@}{0.9ex plus 0.4ex minus 0.2ex}%
{0.3ex plus 0.2ex minus 0ex}{\normalfont\normalsize\itshape}}%
\makeatother

\let\oldthebibliography\thebibliography
\renewcommand{\thebibliography}[1]{%
  \oldthebibliography{#1}%
  \setlength{\itemsep}{-0.2ex}%
  \setlength{\parsep}{0pt}%
}

\author{%
\IEEEauthorblockN{Jiajun Hu\textsuperscript{1}, Ruthwik Reddy Sunketa\textsuperscript{1}, Andrew Boutros\textsuperscript{2}, Aman Arora\textsuperscript{1}}
\IEEEauthorblockA{\textsuperscript{1}Arizona State University, Tempe, AZ, USA \quad \textsuperscript{2}University of Waterloo, Waterloo, Ontario, CA}
\IEEEauthorblockA{\{jiajunh5, rsunketa, aman.kbm\}@asu.edu \quad andrew.boutros@uwaterloo.ca}
}

\begin{document}
%
\title{Boosting FPGA Performance with Direct BRAM-DSP Paths \vspace{-5mm}}

\maketitle

\begin{abstract}
Efficient data movement between memory and compute units is a key performance bottleneck in modern FPGA designs, particularly for deep learning (DL) workloads. In typical FPGA architectures, data transfers between block RAMs (BRAMs) and digital signal processing units (DSPs) must traverse the global routing network, leading to increased wirelength, routing congestion, and critical-path delays. Prior work has explored in- and near-BRAM compute architectures to mitigate these issues, but such solutions often require fundamental changes to FPGA architecture and CAD tools, limiting their commercial viability. This paper proposes a lightweight architectural enhancement that introduces a dedicated direct connection between BRAM and DSP blocks, enabling BRAM data to be consumed by DSPs without passing through the global interconnect. 
We also enhance the placement algorithm to recognize these BRAM–DSP macro blocks. The proposed architectural change incurs negligible area and delay overhead and does not affect non-DL benchmarks, while the proposed CAD remains compatible with the baseline architecture, where it yields negligible change in quality-of-results (QoR).
On an Agilex-10-like FPGA, the proposed architecture and CAD updates deliver up to $+25\%$ Fmax and $-49\%$ wirelength on common DL layer designs.
\end{abstract}


%
\IEEEpeerreviewmaketitle

\section{Introduction}

As modern compute workloads scale significantly, data movement is a growing bottleneck. This impacts FPGAs as well: data must frequently move between various blocks through programmable routing. 
A characteristic feature of contemporary FPGA architectures is the inclusion of dedicated direct interconnects between homogeneous blocks within a column, which can be exploited to reduce reliance on the programmable routing network thereby reducing data movement.
These include cascading BRAMs into deeper logical memories, chaining DSPs for MAC operations and CLB carry chains for larger adders \cite{stratixII,xilinx_ulstrascale}.
However, these hardened paths do not address data movement between heterogeneous blocks, particularly between BRAMs and DSPs, which still traverses the global routing network.

This limitation is acute in Deep Learning (DL) inference, which is dominated by General Matrix--Vector Multiplication (GEMV).
In a GEMV operation, a DSP multiplies a BRAM-resident weight against a runtime-streamed activation through its two input ports. 
Because DL accelerators~\cite{fpga4dl} instantiate thousands of such BRAM-DSP pairs operating in parallel,
the resulting high-volume
BRAM-to-DSP traffic has no direct path, so it traverses the global routing network, creating routing congestion and lengthening the critical path. 
Recent in-/near-BRAM compute architectures cut this data movement but require fundamental changes to the BRAM architecture \& circuitry as well as the CAD toolchain, raising area, programming, and commercial-viability concerns \cite{comefa, bramac}.


To enable efficient BRAM-to-DSP data transfer without disrupting existing FPGA development flows, we propose an enhanced FPGA architecture that has direct paths between BRAMs and DSPs. BRAM output is directly connected to the input of a DSP a few columns away, directly facing this BRAM.
Importantly, the original BRAM and DSP routing interfaces are preserved, maintaining full backward compatibility. 
Since current CAD flows support only  homogeneous (same-type) direct interconnects, they cannot correctly leverage the proposed heterogeneous (cross-type) connections, leading to placement and routing failures.
We enhance the placement engine with simple updates to fully utilize the proposed cross-type paths resulting in no failures and yielding better Quality-of-Results (QoR). 

Our contributions are:
\begin{itemize}
\item \textbf{Architecture.} The first hardened, cross-type direct path between BRAM output ports and DSP inputs, with negligible area and timing overhead and full backward compatibility with default BRAM/DSP interfaces.
\item \textbf{CAD.} An additive placer extension that recognizes cross-type direct connections as placement macros and updates the placement priority of the remaining macros after every successful placement.
\item \textbf{Quantified benefit.} Evaluation shows up to {$+25\%$ Fmax and $-49\%$ wirelength} on DL layer designs, with no degradation on non-DL designs.
\end{itemize}



\section{Background \& Related Work}

\begin{figure}[t]
\centering
\includegraphics[width=\columnwidth]{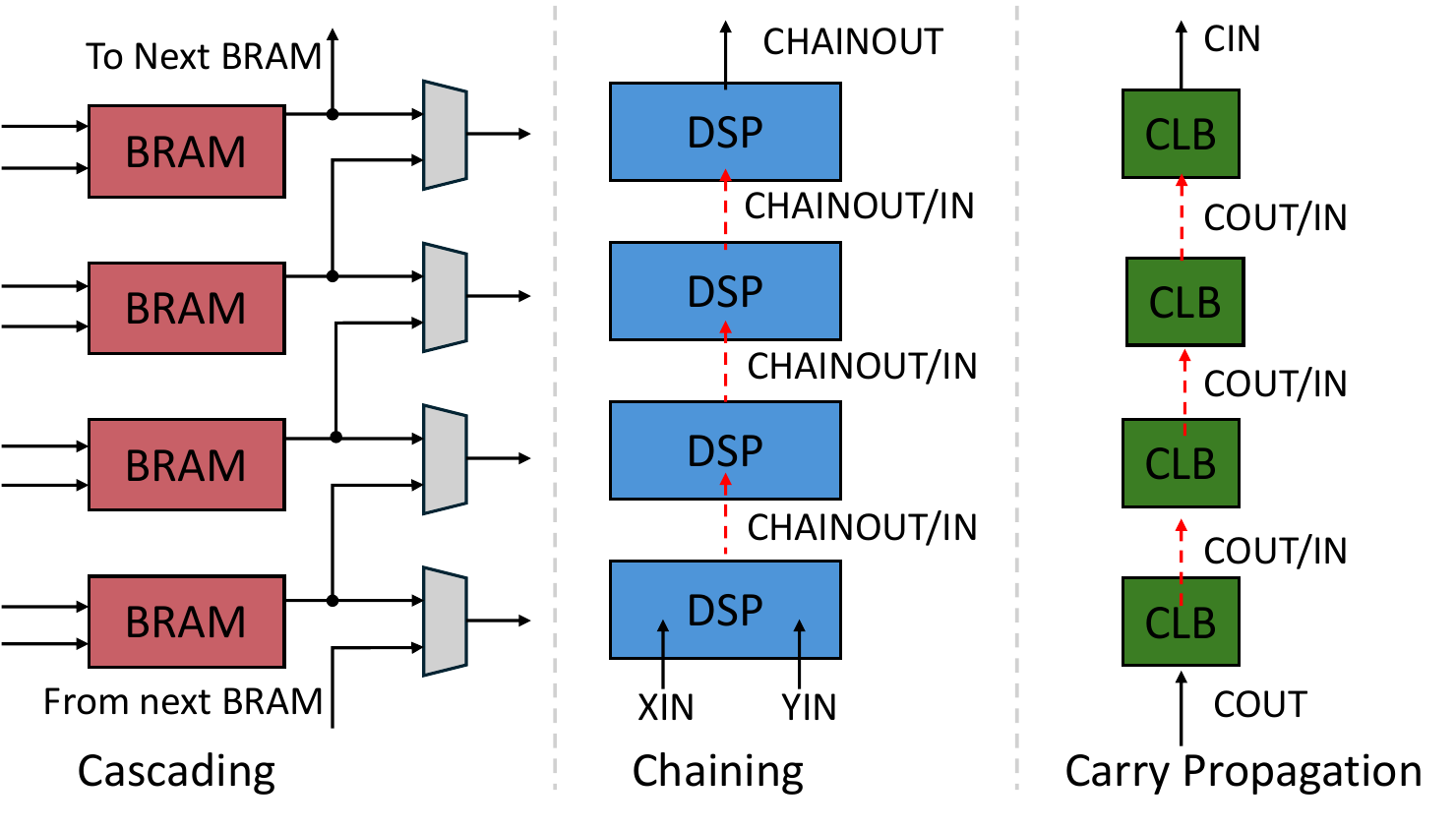}
\caption{\small Example of direct connection between same logic resources in FPGA. BRAM readout is forwarded to upstream. DSPs accept both routed input and direct chaining input. When configured as adders, the carry-bit of upstream CLBs is propagated directly.}
\label{fig:fpga-cascade}
\end{figure}

{\textbf{Same-type Direct Paths.} Modern FPGAs are heterogeneous, tile-based fabrics whose CLB, BRAM, and DSP columns communicate over a flexible but expensive island-style routing network that occupies over half the area and dominates critical-path delay~\cite{fpga_arch}. To sidestep this cost, vendors harden direct interconnects between \emph{same-type} resources within a column (Fig.~\ref{fig:fpga-cascade}): BRAMs cascade into deeper logical memories, DSPs chain to forward partial results, and carry propagation when CLBs are configured as adders. Prior works also leverage these chain optimize their placement to hold high frequency under heavy congestion \cite{RapidLayout,scaling_the_cascade}. However, these changes remain on the same logic resources, we instead harden a BRAM--DSP path to improve efficiency of traffic between different logic resources.}

\textbf{In/near BRAM compute.} Recent proposals push compute into or near the BRAMs and make invasive changes to the BRAM. Some add bit-serial processing elements into the BRAM tile, while other adds a small SRAM buffer and custom ALUs near the BRAMs \cite{comefa,bramac}. Beyond the additional area and power overhead, such architectures need a new programming model and are incompatible with existing FPGA toolchains. Our work instead proposes minimal architecture change that come with full-compatibility to existing CAD flow.


\section{Proposed Architecture \& CAD Changes}
\label{sec:proposed}

\subsection{Proposed Architecture}

\begin{figure*}[tb]
\centering
\includegraphics[width=0.85\linewidth]{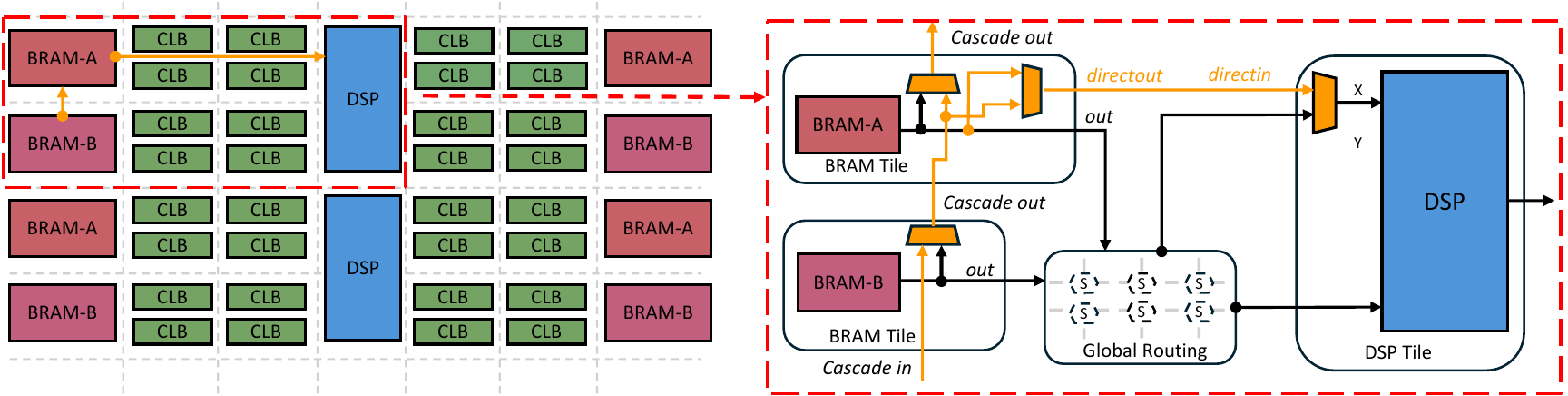}
\caption{\small Final proposed FPGA architecture with two BRAM variants and modified DSPs at BRAM-DSP distance $d = 2$. BRAM-A drives a hardened direct connection to the paired DSP; BRAM-B retains cascade capability only. Both retain their original soft-routed interfaces.}
\label{fig:proposed_arch}
\end{figure*}

At the column level, two BRAM variants, BRAM-A and BRAM-B, are interleaved in each BRAM column, alternating vertically (Fig.~\ref{fig:proposed_arch}, left). Both variants retain full cascade capability and the standard output that drives the general routing, so existing BRAM-DSP communication patterns continue to work unchanged. BRAM-A additionally exposes a \textit{directout} port directly connected to the paired DSP, bypassing the general routing.
At the port level (Fig.~\ref{fig:proposed_arch}, right), a multiplexer (mux) splits BRAM-A's output between the \textit{out} path and \textit{directout}; symmetrically, the DSP's first input is selected between the dedicated \textit{directin} and the standard input. Both muxes reside inside the BRAM and DSP tiles themselves, consuming no routing resources, and are configured per design instance, so an application not using the direct path falls back to the standard datapath with no netlist or programming change.
The horizontal distance $d$ between the BRAM and the DSP is an architectural parameter. Fig.~\ref{fig:proposed_arch} illustrates $d = 2$, but the column interleaving generalizes to any value, which we sweep in our evaluation in Section~\ref{sec:results} to characterize how the direct-path benefit scales with the BRAM--DSP distance.

\subsection{Optimized Placement}

\begin{figure}[t]
\centering
\includegraphics[width=\columnwidth]{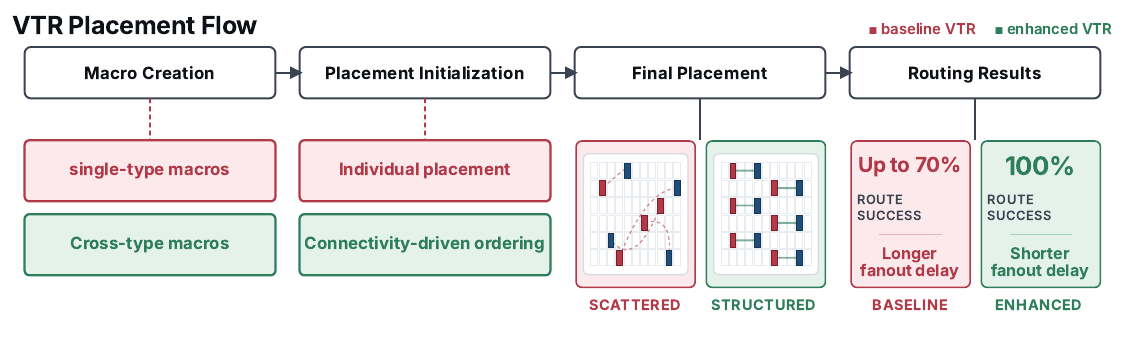}
\caption{\small Proposed VTR CAD changes.}
\label{fig:vtr_flow_mods}
\end{figure}

{The hardened connection only carries a signal when the
paired BRAM and DSP sit at a fixed $(\Delta x, \Delta y)$ offset, where $\Delta x$ is the BRAM--DSP distance.
Baseline VTR places macros individually, with no architecture- or
connectivity awareness that the two must be aligned. Because BRAM and
DSP tiles also occupy different $y$-offsets in the grid, the logical BRAM-DSP pair in the design almost never lands at the required position during placement. The direct path is unroutable on most runs especially for large designs, and the succeeded runs leave the surrounding logic far from BRAM-DSP pair, inflating wirelength and degrading Fmax. We add two extensions, at macro creation and at initial placement
(Fig.~\ref{fig:vtr_flow_mods}), leaving the annealing engine and router untouched, so gains reflect the architecture, not a retuned optimizer.}

\textbf{Cross-type macros.} {We introduce
\textit{cross-type macros} that bind a BRAM-DSP pair or a
BRAM-BRAM-DSP triple when cascading at the
architecturally-mandated $(\Delta x, \Delta y)$ offset during macro
creation. The macro moves as a rigid unit through both initial
placement and simulated annealing, so the required adjacency
is guaranteed for the entire flow.}

\textbf{Connectivity-driven ordering.} VTR places macros in priority order, where a macro's priority rises with $\rho$, its number of already-placed neighbors; so, heavy-connected macros are placed early and their neighbors are tightly clustered. {But VTR refreshes $\rho$ only after a macro is placed by \textit{centroid} heuristic, which itself needs an already-placed neighbor; early placements therefore fall back to random or exhaustive search, $\rho$ stalls, and the initial macros scatter. We refresh $\rho$ after \emph{any} successful placement, weighting each increment by $w(M)$:}
\begin{equation}
w(M)=
\begin{cases}
w_{\max}, & \text{if } M \text{ is a cross-type direct macro,}\\[2pt]
1, & \text{if } \beta\log_2(1{+}\phi(M))+\gamma\log_2(1{+}|M|)\ge T,\\[2pt]
0, & \text{otherwise,}
\end{cases}
\label{eq:c4weight}
\end{equation}
{where $\phi(M)$ is the summed member count of macros $M$ connects to (higher when one is a cross-type macro).
$|M|$ is $M$'s own member count (a cascaded BRAM pair connected to a DSP gives  $|M|=3$ ). $\beta$ and $\gamma$ weight the two log terms and $T$ sets which same-type macros earn the unit increment. A cross-type macro saturates at $w_{\max}$ and is placed before the surrounding region fills. The tuple $(w_{\max},\beta,\gamma,T)$ is exposed as a set of tunable knobs that trade a tighter clustering for cross-type macros and their connected macros against a looser one.Together, the extensions yield a more structured layout, raise routing success on cascade and direct-connection designs.}


\section{Methodology}
\label{sec:methodology}

\textbf{Tools, architecture, and metrics.} 
We use VTR to evaluate the architecture and CAD enhancements.
All experiments use the Agilex-10-like template~\cite{koios} with DSPs in \textit{int\_sop\_4} mode and BRAMs in 512$\times$40 mode, and we report two primary metrics: frequency (Fmax) and total routed wirelength (WL). \textit{Design Compiler}~\cite{synopsysdc} on FreePDK45~\cite{freepdk45} confirms the added multiplexers and direct path are negligible in delay and area relative to the blocks.
For simplicity, we model the direct path as the same wire delay as regular path with no added area.
Precisely modeling the direct path's delay and area would require circuit-level simulation and is left to future work. {The BRAM-DSP distance $d$ is the swept parameter: the main results sweep it while all other experiments fix $d=4$, the architecture's default. All experiments use a fixed grid sized so every benchmark sits at 70-80\% utilization, typical of real deployments, away from both extreme routing congestion. Each configuration reports the average frequency and wirelength over 20 random seeds; because the seed strongly perturbs VPR's initial placement, we assume a $\pm5\%$ band as seed noise and ignore any $\Delta F_\mathrm{max}$ within it.}

\textbf{Benchmarks.} We evaluate the proposed architecture and CAD flow on three types of common DL layers. Table~\ref{tab:dl-benchmarks} reports the resource usage of all six configurations. Each DL layer keeps its weights in BRAM and streams activations from the top module ports. \textit{FC} is a fully-connected layer (GEMV-like, with \textit{ReLU} activation and \textit{LayerNorm}); \textit{Conv} is a single convolution layer with \textit{ReLU} and \textit{BatchNorm}; and \textit{Attention} is from BERT-Tiny with multiple GEMMs and softmax. We choose each layer's dimensions so that it instantiates two weight-storage configurations: a \textit{small} variant whose weights fit in a single $512$-deep BRAM and a \textit{large} variant whose weights require cascaded BRAM pairs. 
Unlike \textit{FC} and \textit{Conv}, where enlarging the weight matrix spills weights into a cascaded BRAM at a fixed DSP count, \textit{Attention}'s projection weights are fixed with model size; its only scalable on-chip memory is the intermediate buffer between GEMM and softmax, which grows with the sequence length $N$. We therefore double $N$ to deepen that buffer into a cascaded chain, reusing the same weights and DSP array. Hence \textit{Attention}'s BRAM grows without simply doubling.

Although the proposed architecture targets DL inference, its minor fabric changes could disturb place-and-route on broader workloads.
To rule this out we evaluate five resource-diverse non-DL VTR-suite designs of varied size and LUT/DSP/BRAM mix: \textit{or1200}, \textit{mkSMAdapter4B}, \textit{bgm}, \textit{LU8PEEng}, and \textit{stereovision2}.

\section{Results}
\label{sec:results}

\subsection{Main Results with BRAM-DSP Distance Sweep}
\label{sec:results:main}

\begin{table}[t]
\centering
\caption{DL benchmark resource usage (CLB shown for the baseline and proposed architectures).}
\label{tab:dl-benchmarks}
\small
\renewcommand{\arraystretch}{0.92}
\begin{tabular}{@{}l r r r r@{}}
\toprule
\textbf{Benchmark} & \textbf{BRAM} & \textbf{DSP} & \multicolumn{2}{c}{\textbf{CLB}} \\
\cmidrule(lr){4-5}
 & & & base & prop \\
\midrule
\textit{FC} (small)        & 256 & 256 & 1710 & 1710 \\
\textit{FC} (large)        & 512 & 256 & 2039 & 1706 \\
\textit{Conv} (small)      & 256 & 260 & 1753 & 1753 \\
\textit{Conv} (large)      & 512 & 260 & 2271 & 1755 \\
\textit{Attention} (small) & 320 & 323 & 1546 & 1546 \\
\textit{Attention} (large) & 450 & 323 & 1863 & 1547 \\
\bottomrule
\end{tabular}
\end{table}

\begin{figure}
    \centering
    \includegraphics[width=\columnwidth]{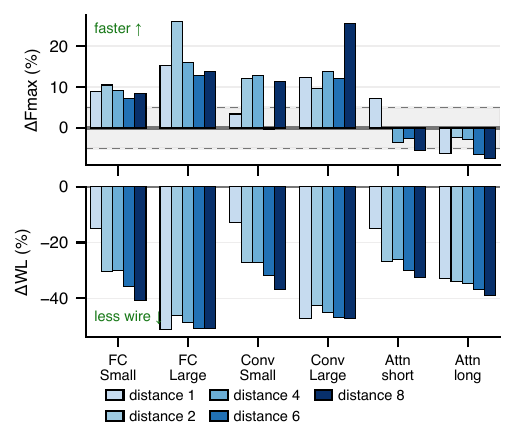}
    \caption{Frequency and wirelength improvements of the proposed system over the baseline (default VTR), across six DL kernels and BRAM--DSP distances.}
    \label{fig:main_sa}
\end{figure}

Fig.~\ref{fig:main_sa} compares the {proposed system (proposed architecture $+$ proposed CAD) against the baseline system (baseline architecture $+$ default VTR)}.
Wirelength is {consistently reduced, by up to $\sim$49\%} as expected, but frequency gains are not uniform: the critical path depends on each design's RTL, and the BRAM-DSP connection is never itself critical in any of the six benchmarks, the soft logic is. \textit{FC} and \textit{Conv} run only lightweight activation and address generation, so their critical path is the placement-sensitive address-generator fanout, which the direct path's tighter placement shortens up to $+25\%$. \textit{Attention} is instead bound by the \textit{softmax} reduction tree and \textit{LayerNorm} variance computation, which the direct path cannot touch, so its frequency stays flat.
Larger benchmarks are expected to achieve similar wirelength reductions; frequency gains will depend on  the critical path.
However, the proposed architecture offers a broader benefit by reducing pressure on the global routing fabric, potentially improving overall routability and timing closure across the design. Investigating these effects on larger benchmarks such as \textit{Koios} \cite{koios} and \textit{Titan} is future work \cite{ titan}.

\subsection{Robustness}
\label{sec:results:scalability}
\begin{figure}
  \centering
  \includegraphics[width=\columnwidth]{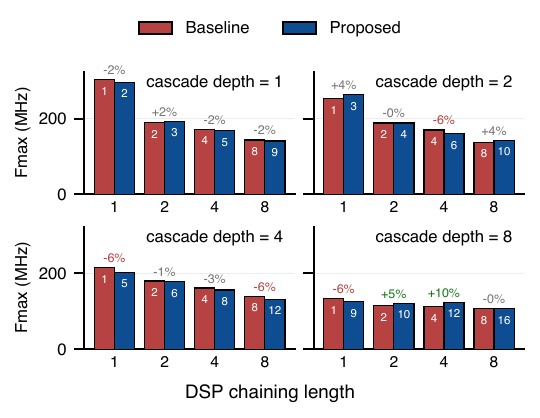}
  \caption{DSP chain length $\times$ BRAM cascade depth robustness sweep at distance of 4.}
  \label{fig:scalability}
\end{figure}


In addition to DSP chains and BRAM cascades, the proposed direct BRAM--DSP connection enables larger \emph{heterogeneous} macros, e.g., a U-shaped macro with multiple cascaded BRAMs in one column, a BRAM-DSP direct link, and a DSP chain in another column. Such macros are harder to relocate during simulated annealing, since they {induce much larger} placement perturbations.
Fig.~\ref{fig:scalability} sweeps DSP-chain length and BRAM-cascade depth on a synthetic benchmark exercising the direct connection; the baseline is the same architecture under default VTR. We observe that even as the fused macro grows to the largest cascade and chaining size, it stays placeable across the entire grid and {frequency stays similar to the baseline}, imposing no placement penalty under high utilization.

\subsection{Do-No-Harm}
\label{sec:results:noharm}

\begin{figure}
  \centering
  \includegraphics[width=\columnwidth]{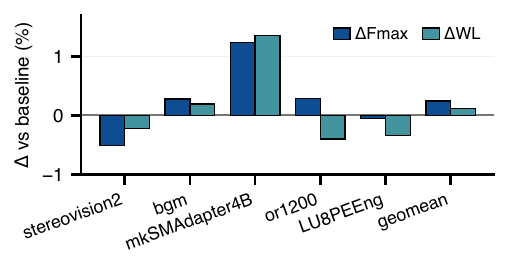}
  \caption{The proposed architecture on the non-DL benchmarks.}
  \label{fig:donoharm}
\end{figure}
{To better assess the cost of proposed architecture and CAD changes, we evaluate the frequency and wirelength of proposed architecture in non-DL VTR benchmarks, as well as the runtime and memory footprint of proposed CAD. Fig.~\ref{fig:donoharm} shows the proposed architecture leaves both frequency and wirelength essentially unchanged from the baseline, preserving the FPGA's general-purpose flexibility. For CAD, running the proposed and the baseline CAD on the same idle server across all benchmarks yields nearly identical runtime and memory footprint (plots omitted for space).}


\section{Conclusion}
\label{sec:Conclusion}
We present a lightweight FPGA enhancement: a hardened cross-type BRAM--DSP direct path and an additive placer extension for the resulting macros, fully compatible with existing FPGA CAD flows. On an Agilex-10-like template it delivers up to $+25\%$ Fmax and $-49\%$ wirelength on DL layers, while leaving non-DL benchmarks and the baseline CAD flow performance essentially unchanged. Future work will involve evaluating the proposed architecture and CAD changes using larger benchmark suites such as \textit{Koios} \cite{koios} and \textit{Titan} \cite{titan}.



%

\bibliographystyle{ieeetr}
\bibliography{bibfile}



\end{document}